\documentclass{article} 

\usepackage{latexsym}
\usepackage{amsmath, amssymb}
\usepackage{amsthm}
\usepackage{amsmath}
\usepackage{graphicx}

\newtheorem{thm}{Theorem}[section]

\newtheorem{cor}{Corollary}[section]
\newtheorem{pro}{Proposition}[section]

\newtheorem{rem}{Remark}[section]

\newcommand{\CM}{\mathbb{C}} 

\newcommand{\ZM}{\mathbb{Z}}

\newcommand{\PM}{\mathbb{P}}

\title{{\Large {\bf Zeta function for a fuzzy cellular automaton}}
\author{
{\small Jiro Akahori}\\
{\scriptsize Department of Mathematical Sciences, 
Ritsumeikan University}\\
{\scriptsize 1-1-1 Noji-higashi, Kusatsu, 525-8577, Japan}\\
{\scriptsize e-mail: akahori@se.ritsumei.ac.jp}\\
{\small Norio Konno}\\
{\scriptsize Department of Mathematical Sciences, 
Ritsumeikan University}\\
{\scriptsize 1-1-1 Noji-higashi, Kusatsu, 525-8577, Japan}\\
{\scriptsize e-mail: n-konno@fc.ritsumei.ac.jp}\\
{\small Iwao Sato}\\
{\scriptsize Oyama National College of Technology}\\
{\scriptsize Oyama, Tochigi 323-0806, Japan}\\
{\scriptsize e-mail: satoz92711@gmail.com} \\
{\small Sennosuke Watanabe} \\
{\scriptsize Faculty of Informatics, The University of Fukuchiyama} \\
{\scriptsize 3370 Azahori, Fukuchiyama, Kyoto 620-0886, Japan}\\
{\scriptsize e-mail: sewatana@fukuchiyama.ac.jp} \\
{\small Xinmiao Zhang }\\
{\scriptsize Department of Mathematical Sciences, Ritsumeikan University}\\
{\scriptsize 1-1-1 Noji-higashi, Kusatsu, 525-8577, Japan}\\
{\scriptsize e-mail: ra0108ki@ed.ritsumei.ac.jp}}
}
\date{\empty }
\begin{document}
\maketitle

\par\noindent
\begin{small}
\par\noindent
{\bf Abstract}. We introduce a fuzzification of Rule 90 as an interacting particle system (IPS) 
and define the associated IPS-type zeta function by using its global operator.
For this fuzzy Rule 90, we derive a determinant factorization based on a
Hadamard block decomposition and obtain a recursive determinant formula.
We also give an explicit trace formula for the global operator and compute
the zeta functions explicitly for \(N=2,3,4,5\).
Furthermore, in the deterministic case \(v=0\), namely the ordinary Rule 90,
we prove a dyadic product formula for the zeta function for all \(N\geq2\).

{\it Abbr. title:} The zeta function for a fuzzy cellular automaton



\end{small}

\section{Introduction}
As a quantum counterpart of the classical random walk, the quantum walk has recently attracted considerable attention in various fields. 
For reviews and books on quantum walks, see, for example, Ambainis \cite{Ambainis2003}, Kempe \cite{Kempe2003}, Kendon \cite{Kendon2007}, 
Konno \cite{Konno2008b}, Venegas-Andraca \cite{VA}, Manouchehri and Wang \cite{MW}, Portugal \cite{Portugal2013}. 

In our previous paper \cite{KKS1}, Komatsu et al. studied a relation between the Grover walk and the zeta function based on 
the Konno-Sato theorem \cite{KS2011} 
and called this relation ``Grover/Zeta Correspondence". 
Furthermore, in the subsequent paper \cite{KKS2}, Komatsu et al. extended the Grover walk with F-type and the positive-support version 
of the Grover walk with F-type to a class of walks with both F- and M-types by using not the Konno-Sato theorem but a method of 
the Fourier transform for the case of the $d$-dimensional torus. Our class contains QWs and CRWs. 
In addition, we treated the open quantum random walk (OQRW) which has the CRW as a special model. 

The Grover walk is one of the most well-investigated quantum walks (QWs) inspired by the famous Grover algorithm \cite{Grover}. 
The QW is a quantum counterpart of the correlated random walk (CRW) which has the random walk (RW) as a special model. 
In fact, the CRW is the RW with memory. 

Our previous two papers \cite{KKS1,KKS2} investigated one-particle models including RWs, CRWs, QWs, and OQRWs. 
Komatsu et al. \cite{KKS3} dealt with  multi-particle models with probabilistic or quantum interactions, called the {\em interacting particle systems} (IPS) 
including probabilistic cellular automata (PCA) and quantum cellular automata (QCA). In particular, PCA have oriented percolation as a special model. 
We introduced the IPS-type zeta function and called ``IPS/Zeta Correspondence". 

Cellular automata (CA) are discrete dynamical system consisting of cell configurations that 
evolve according to local update rule based on neighboring cells.
Simple CA are 2-state 3-neighbor CA by Wolfram \cite{Wolfram}, called elementary CA (ECA), 
with the state set $\{0,1\}$.
ECA Rule 90 is known as one of the interesting ECA that generates a fractal figure called a Sierpinski gasket, 
and various applications of this rule have been studied so far \cite{Sakai,Wolfram}.
\par
The fuzzification introduced in \cite{Cattaneo} is a scheme to obtain continuous-valued CA from ECA.
Such CA are called fuzzy ECA and their local rules are polynomials corresponding to the states and the neighbors.
The asymptotic behaviors of fuzzy ECA are studied for cases with a single seed in a zero background \cite{Mingarelli1} 
and for periodic cases \cite{Betel}.
There are more detailed studies on dynamics for some specific rules, e.g., Rule 90 \cite{Yacoubi,Flocchini},
Rule 110 \cite{Mingarelli2}, Rule 184 \cite{Mingarelli3}.
In recent years, Nishida et al. proposed a fuzzification method for general CA \cite{Nishida1}, 
and their application to traffic flow models are also investigated \cite{Nishida2,Nishida3}.

The rest of the present manuscript is organized as follows. 
In Section 2, we give a survey for the IPS with $N$ sites and its IPS zeta function. 
In Section 3, we introduce a fuzzification of Rule 90, and define the matrix zeta function with respect to its global operator. 
Furthermore, we give an explicit formula for the matrix zeta function of a fuzzification of Rule 90 with $N=2$. 
In Section 4, we present a formula for the trace of the global operator of a fuzzification of Rule 90. 
In Section 5, we treat the matrix zeta function of a fuzzification of Rule 90 for a general $N$. 
In Section 6, we compute certain determinants. 
In Section 7, we give an explicit formula for the matrix zeta function of a fuzzification of Rule 90 with $N=3$. 
In Section 8, we derive a determinant factorization based on Hadamard block decomposition and compute the zeta functions explicitly 
for $N=4,5$. 
In Section 9, we prove a dyadic product formula for the zeta function of Rule 90 for all $N \geq 2$. We should remark that, in Komatsu et al.~\cite{KKS3},
this formula was established for \(N=2,3,4\),  was stated as a
conjecture for general \(N\).
Therefore, our result implies that their conjecture is correct.

\section{IPS zeta function}

This section introduces our models, i.e., IPSs. 
To do so, we first introduce the following notation: $\mathbb{Z}$ is the set of integers, $\mathbb{Z}_{\ge}$ is the set of non-negative integers, 
$\mathbb{Z}_{>}$ is the set of positive integers, $\mathbb{R}$ is the set of real numbers, and $\mathbb{C}$ is the set of complex numbers. 


Let $\mathbb{P}_N = \{ 0,1, \ldots, N-1 \}$ be the site space with $N$ sites. 
Throughout this paper, we mainly assume that $N \ge 2$. There are two states ``0" or ``1" at each site. 
Let $\eta (x) \in \{0,1\}$ denote the state of the site $x \in \PM_N$, i.e., $x=0,1,\ldots, N-1$. Note that $\eta(x)$ is also denoted by $\eta_x$.
The configuration space is $\{0,1\}^{\mathbb{P}_N}$ with $2^N$ elements. 
A configuration $\eta = (\eta (0), \eta (1), \ldots, \eta (N-1)) \in \{0,1\}^{\PM_N}$ can be interpreted in terms of particle occupation as follows: 
$\eta (x)=1$ means that a particle exists at site $x \in \PM_N$, and $\eta (x)=0$ means that $x$ is vacant. 
In this paper, we put
\begin{align}
|0\rangle =
\begin{bmatrix}
1 \\
0
\end{bmatrix},
\quad
|1\rangle =
\begin{bmatrix}
0 \\
1
\end{bmatrix}.
\end{align}

For example, when $N=3$, a configuration $(0,0,1) \in \{0,1\}^{\mathbb{P}_3}$ means that the state ``0" at site 0, the state ``0" at site 1, 
and the state ``1" at site 2. In other words, $(\eta (0), \eta (1), \eta (2)) = (0,0,1)$. 
We also write $(0,0,1)$ as
\[
|0\rangle\,|0\rangle\,|1\rangle
\;=\;
|0\rangle\otimes|0\rangle\otimes|1\rangle,
\]
where $\otimes$ denotes the tensor product.

To define our model, we introduce the {\em local} operator $Q^{(l)}$ and the {\em global} operator $Q^{(g)}_N$ in the following way. 
This definition is based on Katori et al. \cite{KatoriEtAl2004}.

We first define the $4 \times 4$ matrix $Q^{(l)}$ by
\[
Q^{(l)}
=\left[ 
\begin{array}{cccc}
a^{00}_{00} & a^{01}_{00} & a^{10}_{00} & a^{11}_{00} \\ 
a^{00}_{01} & a^{01}_{01} & a^{10}_{01} & a^{11}_{01} \\ 
a^{00}_{10} & a^{01}_{10} & a^{10}_{10} & a^{11}_{10} \\ 
a^{00}_{11} & a^{01}_{11} & a^{10}_{11} & a^{11}_{11}  
\end{array}  
\right] 
,
\] 
where $a^{ij}_{kl} \in \mathbb{C}$ for $i,j,k,l \in \{0,1\}.$  Let $\eta_n (x) \in \{0,1\}$ denote the state of the site $x \in \PM_N$ at time $n \in \ZM_{\ge}$. 
The element of $Q^{(l)}$, $a^{ij}_{kl}$, means the {\em transition weight} from $(\eta_n (x), \eta_n (x+1)) =(i,j)$ to $(\eta_{n+1} (x), \eta_{n+1} (x+1))=(k,l)$ 
for any $x=0,1, \ldots, N-2$ and $n \in \ZM_{\ge}$. If $a^{ij}_{kl} \in [0,1]$, then the transition weight can be interpreted as a transition probability. 
We call ``$x$" the {\em left site} and ``$x+1$" the {\em right site}. 
Throughout this paper, we assume that $a^{ij}_{kl}=0$ if $j \not=l$. In other words, after the time transition, the state of the right site does not change. 
This assumption is necessary to define the global operator $Q^{(g)}_N$ described below. Therefore, under this assumption, $Q^{(l)}$ is rewritten as
\[ 
Q^{(l)}
=\left[ 
\begin{array}{cccc}
a^{00}_{00} & \cdot & a^{10}_{00} & \cdot \\ 
\cdot & a^{01}_{01} & \cdot & a^{11}_{01} \\ 
a^{00}_{10} & \cdot & a^{10}_{10} & \cdot \\ 
\cdot & a^{01}_{11} & \cdot & a^{11}_{11} 
\end{array}  
\right] 
, 
\]
where ``$\cdot$" means 0. By definition, the interaction of our IPS is nearest neighbor. In particular, if $a^{ij}_{kl} \in \{0,1\}$, 
then the IPS is called the {\em cellular automaton} (CA). Next we define the $2^N \times 2^N$ matrix $Q^{(g)}_N$ by
\begin{align*}
Q^{(g)}_N
&= \left( I_2 \otimes I_2 \otimes \cdots \otimes I_2 \otimes Q^{(l)} \right) \left( I_2 \otimes I_2 \otimes \cdots \otimes Q^{(l)} \otimes I_2 \right) 
\\
& \qquad \cdots \left( I_2 \otimes Q^{(l)} \otimes \cdots \otimes I_2 \otimes I_2 \right) \left( Q^{(l)} \otimes I_2 \otimes \cdots \otimes I_2 \otimes I_2 \right),
\end{align*}
where $I_n$ is the $n \times n$ identity matrix.

We now specify the conditions on the local operator \(Q^{(l)}\)
for the two classes mentioned in Section~1: PCA and QCA.
PCA are also known as stochastic cellular automata.
For PCA, the local transition weights satisfy
\begin{align}
a^{00}_{00} + a^{00}_{10} = a^{01}_{01} + a^{01}_{11} = a^{10}_{00} + a^{10}_{10} = a^{11}_{01} + a^{11}_{11} = 1, \quad a^{ij}_{kj} \in [0,1].
\label{condPCA}
\end{align}
That is, $Q^{(l)}$ becomes a transposed {\em stochastic matrix} (also called {\em transition matrix}). 
Furthermore, we easily see that `` $Q^{(l)}$ is a transposed stochastic matrix if and only if $Q^{(g)}_N$ is a transposed stochastic matrix". 
In other words, the sum of the elements of any column for $Q^{(l)}$ or $Q^{(g)}_N$ is equal to 1.

On the other hand, a QCA model is characterized by the unitarity of $Q^{(l)}$ , i.e., 
\begin{align*}
&
|a^{00}_{00}|^2 + |a^{00}_{10}|^2 = |a^{01}_{01}|^2 + |a^{01}_{11}|^2 = |a^{10}_{00}|^2 + |a^{10}_{10}|^2 = |a^{11}_{01}|^2 + |a^{11}_{11}|^2 = 1,
\nonumber
\\
& 
\qquad a^{00}_{00} \  \overline{a^{10}_{00}} + a^{00}_{10} \  \overline{a^{10}_{10}} =
 a^{01}_{01} \  \overline{a^{11}_{01}} + a^{01}_{11} \  \overline{a^{11}_{11}} =0. 
\end{align*}
This QCA was introduced by Konno~\cite{Konno2008b} as a quantum counterpart
of the Domany--Kinzel model~\cite{DomanyKinzel1984}.  As in the case of PCA,
we easily see that \(Q^{(l)}\) is unitary if and only if \(Q_N^{(g)}\) is
unitary.
The evolution of IPS on $\PM_N$ is determined by 
\begin{align*}
\eta_{n} = \left( Q^{(g)} _N \right)^n \eta_{0} \quad (n \in \ZM_{\ge})
\end{align*}
for an initial state $\eta_{0}$. Note that $\eta_{n}, \eta_{0} \in \CM^{2^N}$ and $Q^{(g)} _N$ is a $2^N \times 2^N$ matrix.

We give some typical models.  
The Domany-Kinzel (DK) model in PCA is defined by
\[
Q^{(l)}_{DK}
=\left[ 
\begin{array}{cccc}
1 & \cdot & 1-p & \cdot \\ 
\cdot & 1-p & \cdot & 1-q \\ 
\cdot & \cdot & p & \cdot \\ 
\cdot & p & \cdot & q   
\end{array}  
\right] 
,
\]
where $p, \ q \in [0,1]$. In particular, if $q=p$ (resp. $q=1-(1-p)^2$), then the DK model becomes oriented site (resp. bond) percolation 
(see Durrett~\cite{Durrett1988} and Katori et al.~\cite{KatoriEtAl2004}, for example).

Next, the well-known Wolfram {\em Rule 90} (see \cite{Wolfram}) is 
\[
Q^{(l)}_{90} 
=\left[ 
\begin{array}{cccc}
1 & \cdot & \cdot & \cdot \\ 
\cdot & \cdot & \cdot & 1 \\ 
\cdot & \cdot & 1 & \cdot \\ 
\cdot & 1 & \cdot & \cdot   
\end{array}  
\right] 
.
\]

Now, we state the IPS zeta function.
Following the walk-type zeta function studied in our previous papers
\cite{KKS1,KKS2}, we define the \emph{IPS-type zeta function} by
\begin{align}
\overline{\zeta} \left(Q^{(l)}, \PM_N, u \right) = \det 
\left\{
\Big( I_{2^N} - u Q^{(g)}_N \Big)
\right\}
^{-1/2^N} \quad (N \in \ZM_{>}).
\end{align}
Note that when $N=1$, we put $Q^{(g)}_1 = I_2$. Thus we have 
\begin{align}
\overline{\zeta} \left(Q^{(l)}, \PM_1, u \right) = (1-u)^{-1}.
\end{align}
The difference between the walk-type and IPS-type zeta functions is that the former is based on the one-particle on lattice space and 
the latter is based on multi-particle on configuration space. Concerning the zeta function, see \cite{KKS1,KS2011}, for example.

Furthermore, we define \(C_r(Q^{(l)},\PM_N)\) by
\begin{align}
\overline{\zeta} \left(Q^{(l)},\PM_N,u\right)
=
\exp\left(
\sum_{r=1}^{\infty}
\frac{C_r(Q^{(l)},\PM_N)}{r}u^r
\right)
\qquad (N\in\ZM_{>}).
\label{CrDef}
\end{align}
Note that, when \(N=1\), Eq. (4) implies
\begin{align}
C_r (Q^{(l)}, \PM_1) = 1 \quad (r \in \ZM_{>}).
\end{align}
We note that $C_r (Q^{(l)}, \mathbb{P}_N)$ corresponds to a kind of return rate of $2^N$ configurations at time $r$. 
For example, in the case of PCA, if $C_r (Q^{(l)}, \mathbb{P}_N) = 1$, then $( Q^{(g)}_N )^r = I_{2^N}$, i.e., the period is $r$. Then the following result holds. 
We give the proof of it. 


\begin{pro}[Komatsu, Konno, Sato \cite{{KKS3}}] \label{prp1}
\begin{align*}
C_r (Q^{(l)}, \PM_N) = \frac{1}{2^N} {\rm Tr} \left( \left( Q^{(g)}_N \right)^r \right) \quad (r, N \in \ZM_{>}).
\end{align*}
\end{pro} 
\begin{proof}
For any square matrix $\mathbf A$, we have
\begin{align*}
    \det (I_n - u\mathbf A )&= 
    \exp \left\{ \log (\det(I_n - u\mathbf A))\right\} \\
    &= \exp \left\{ {\rm Tr} (\log ( I_n - u\mathbf A))\right\},
\end{align*}
and 
\begin{align*}
    \log (I_n - u\mathbf A)= - \sum_{r=1}^{\infty} \frac{(u\mathbf A)^r}{r} .
\end{align*}
Thus, we have 
\begin{align*}
    {\rm Tr} (\log (I_n - u\mathbf A)) = - \sum_{r=1}^{\infty} \frac{u^r}{r}  {\rm Tr} (\mathbf A^r).
\end{align*}
When $n= 2^N$ and $ \mathbf A = Q_N^{(g)}$ , 
\begin{align*}
    \det (I_{2^N} - uQ_N^{(g)}) = \exp \left\{ - \sum_{r=1}^{\infty} \frac{u^r}{r} {\rm Tr} ( (Q_N^{(g)})^r ) \right\},
\end{align*}
The IPS-type Zeta function can be written by
\begin{align*}
     \overline{\zeta} \left(Q^{(l)}, \PM_N, u \right) = & \left\{ \det(I_{2^N} - uQ_N^{(g)}) \right\}^{- 1/2^N} \\
     & = \exp \left\{ - \frac{1}{2^N} \log ( \det(I_{2^N}- uQ_N^{(g)}))\right\} \\
     & = \exp \left\{ \frac{1}{2^N}\sum_{r=1}^{\infty} \frac{u^r}{r} {\rm Tr} ((Q_N^{(g)})^r)\right\} \\
     & = \exp \left\{ \sum_{r=1}^{\infty} \frac{Cr(Q^{(l)},\PM_N)}{r} u^r \right\}.
\end{align*}
Hence,
\begin{align*}
    C_r (Q^{(l)}, \PM_N) = \frac{1}{2^N} {\rm Tr} \left( \left( Q^{(g)}_N \right)^r \right) . 
\end{align*}
\end{proof}

\section{Fuzzification of Rule 90}

The local operator of Rule 90 is  
\[
Q^{(l)}_{90} 
=\left[ 
\begin{array}{cccc}
1 & \cdot & \cdot & \cdot \\ 
\cdot & \cdot & \cdot & 1 \\ 
\cdot & \cdot & 1 & \cdot \\ 
\cdot & 1 & \cdot & \cdot   
\end{array}  
\right] 
.
\]
Then the local operator of the {\em fuzzification} of Rule 90 (or {\em fuzzy Rule 90}) is defined as follows:  
\[
Q^{(l)}_{f} 
=\left[ 
\begin{array}{cccc}
1-v & v & v & v \\ 
v & v & v & 1-v \\ 
v & v & 1-v & v \\ 
v & 1-v & v & v   
\end{array}  
\right] 
,
\]
where $0 \leq v \leq 1$. If $v=0$, then $Q^{(l)}_{f}$ becomes $Q^{(l)}_{90}$. Thus, our fuzzification of Rule 90 contains Rule 90 as a special case. Furthermore, the global operator of the fuzzification of Rule 90 is the  $2^N \times 2^N$ matrix $Q^{(g)}_{N,f}$ defined as follows:
\begin{align*}
Q^{(g)}_{N,f}
&= \left( I_2 \otimes I_2 \otimes \cdots \otimes I_2 \otimes Q^{(l)}_f \right) \left( I_2 \otimes I_2 \otimes \cdots \otimes Q^{(l)}_f \otimes I_2 \right) 
\\
& \qquad \cdots \left( I_2 \otimes Q^{(l)}_f \otimes \cdots \otimes I_2 \otimes I_2 \right) \left( Q^{(l)}_f \otimes I_2 \otimes \cdots \otimes I_2 \otimes I_2 \right) .
\end{align*}
Moreover, we introduce a zeta function for the fuzzification of Rule 90 as follows: 
\[
\overline{\zeta } (Q^{(l)}_f , \PM_N ,u)= \det \left\{ ( I_{2^N} -u Q^{(g)}_{N,f} ) \right\}^{-1/2^N}.
\]
If $N=1$, then we have 
\[
\overline{\zeta } (Q^{(l)}_f , \PM_1 ,u)= \left\{ \det ( I_2 -u I_2 ) \right\}^{-1/2} = \{ (1-u)^2 \}^{-1/2} =(1-u)^{-1} . 
\]

Now, we consider the result for $\overline{\zeta } (Q^{(l)}_f , \PM_2 ,u)$.

\begin{thm} \label{thm3}
\begin{align*}
\overline{\zeta } (Q^{(l)}_f , \PM_2 ,u)^{-1}=(1-(1-2v)u)^{1/2} (1-(1+2v)u)^{1/4} (1-(-1+2v)u)^{1/4} . 
\end{align*}
\end{thm} 

\begin{proof}
Since 
\[
Q^{(g)}_{2,f} = Q^{(l)}_f , 
\]
we have 
\begin{align*}
\det\!\bigl(\lambda I_{2^2}-Q^{(g)}_{2,f}\bigr)
&=
\det
\begin{bmatrix}
\lambda-1+v & -v & -v & -v \\
-v & \lambda-v & -v & -1+v \\
-v & -v & \lambda-1+v & -v \\
-v & -1+v & -v & \lambda-v
\end{bmatrix}
\\[10pt]
&=
\det
\begin{bmatrix}
\lambda-1-2v & -v & -v & -v \\
\lambda-1-2v & \lambda-v & -v & -1+v \\
\lambda-1-2v & -v & \lambda-1+v & -v \\
\lambda-1-2v & -1+v & -v & \lambda-v
\end{bmatrix}
\\[10pt]
&=
(\lambda-1-2v)\det
\begin{bmatrix}
1 & -v & -v & -v \\
1 & \lambda-v & -v & -1+v \\
1 & -v & \lambda-1+v & -v \\
1 & -1+v & -v & \lambda-v
\end{bmatrix}
\\[10pt]
&=
(\lambda-1-2v)\det
\begin{bmatrix}
1 & -v & -v & -v \\
0 & \lambda & 0 & 2v-1 \\
0 & 0 & \lambda-1+2v & 0 \\
0 & 2v-1 & 0 & \lambda
\end{bmatrix}
\\[10pt]
&=
(\lambda-1-2v)\det
\begin{bmatrix}
\lambda & 0 & 2v-1 \\
0 & \lambda-1+2v & 0 \\
2v-1 & 0 & \lambda
\end{bmatrix}
\\[10pt]
&=
(\lambda-1-2v)
\bigl\{
\lambda^2(\lambda-1+2v)
-(2v-1)^2(\lambda-1+2v)
\bigr\}
\\[8pt]
&=
(\lambda-1-2v)(\lambda-1+2v)
\bigl(\lambda^2-(2v-1)^2\bigr)
\\[8pt]
&=
(\lambda-1-2v)(\lambda-1+2v)^2(\lambda+1-2v).
\end{align*}
Thus, 
\[
{\rm Spec} (Q^{(g)}_{2,f})= \{ 1-2v, 1-2v, 1+2v, -1+2v \} , 
\]
where ${\rm Spec} ( \mathbf F )$ is the set of all eigenvalues of a square matrix $\mathbf F $. 
Therefore, we obtain 
\[
{\rm Tr} (Q^{(g)}_{2,f})=2(1-2v)+(1+2v)+(-1+2v)=2. 
\]
Furthermore, we have 
\[
{\rm Tr} ((Q^{(g)}_{2,f})^r )=2 (1-2v)^r +(1+2v)^r +(-1+2v)^r . 
\]
Thus, 
\begin{align*}
&\log\!\left(
\overline{\zeta}\!\left(Q_f^{(l)},\PM_2,u\right)^{-1}
\right)
\\[8pt]
&=
-\frac{1}{2^2}
\left\{
2\sum_{r=1}^{\infty}\frac{(1-2v)^r}{r}u^r
+\sum_{r=1}^{\infty}\frac{(1+2v)^r}{r}u^r
+\sum_{r=1}^{\infty}\frac{(-1+2v)^r}{r}u^r
\right\}
\\[8pt]
&=
\frac{1}{2}\log\!\bigl(1-(1-2v)u\bigr)
+\frac{1}{4}\log\!\bigl(1-(1+2v)u\bigr)
+\frac{1}{4}\log\!\bigl(1-(-1+2v)u\bigr).
\end{align*}
Hence, 
\[
\overline{\zeta } (Q^{(l)}_f , \PM_2 ,u)^{-1}=(1-(1-2v)u)^{1/2} (1-(1+2v)u)^{1/4} (1-(-1+2v)u)^{1/4} . 
\]
\end{proof}

\section{Trace of global operator with respect to fuzzification of Rule 90} 

Although determining the eigenvalues of the global operator of fuzzy Rule 90 is difficult, its trace can be calculated explicitly.

The trace of the global operator of the fuzzification of Rule 90 is given as follows.

\begin{thm} \label{thm4}
\begin{align*}
{\rm Tr} (Q^{(g)}_{N,f} )=2(1+2v)^{N-2} \quad (N \geq 2) . 
\end{align*}
\end{thm} 

\begin{proof}
At first, set $Q^{(g)}_{N} = Q^{(g)}_{N,f} $. 
Furthermore, let 
\[
Q^{(g)}_{N} = \left[ 
\begin{array}{cc}
{\bf A}_N & {\bf B}_N \\ 
{\bf C}_N & {\bf D}_N 
\end{array}  
\right] 
. 
\]
Remark that $Q^{(g)}_N$ is a $2^N \times 2^N$ matrix, and
$\mathbf A_N$, $\mathbf B_N$, $\mathbf C_N$, and $\mathbf D_N$ are
$2^{N-1} \times 2^{N-1}$ matrices.
Since 
\[
Q^{(g)}_{N+1} =( I_2 \otimes Q^{(g)}_{N} )(Q^{(l)} \otimes I_{2^{N-1}} ) , 
\]
we have 
\begin{align*}
&Q^{(g)}_{N+1}
\\[8pt]
&=
\begin{bmatrix}
\mathbf{A}_N & \mathbf{B}_N & \mathbf{0} & \mathbf{0} \\
\mathbf{C}_N & \mathbf{D}_N & \mathbf{0} & \mathbf{0} \\
\mathbf{0} & \mathbf{0} & \mathbf{A}_N & \mathbf{B}_N \\
\mathbf{0} & \mathbf{0} & \mathbf{C}_N & \mathbf{D}_N
\end{bmatrix}
\begin{bmatrix}
(1-v)\mathbf{I} & v\mathbf{I} & v\mathbf{I} & v\mathbf{I} \\
v\mathbf{I} & v\mathbf{I} & v\mathbf{I} & (1-v)\mathbf{I} \\
v\mathbf{I} & v\mathbf{I} & (1-v)\mathbf{I} & v\mathbf{I} \\
v\mathbf{I} & (1-v)\mathbf{I} & v\mathbf{I} & v\mathbf{I}
\end{bmatrix}
\\[10pt]
&=
\begin{bmatrix}
(1-v)\mathbf{A}_N+v\mathbf{B}_N
&
v\mathbf{A}_N+v\mathbf{B}_N
&
v\mathbf{A}_N+v\mathbf{B}_N
&
v\mathbf{A}_N+(1-v)\mathbf{B}_N
\\[4pt]
(1-v)\mathbf{C}_N+v\mathbf{D}_N
&
v\mathbf{C}_N+v\mathbf{D}_N
&
v\mathbf{C}_N+v\mathbf{D}_N
&
v\mathbf{C}_N+(1-v)\mathbf{D}_N
\\[4pt]
v\mathbf{A}_N+v\mathbf{B}_N
&
v\mathbf{A}_N+(1-v)\mathbf{B}_N
&
(1-v)\mathbf{A}_N+v\mathbf{B}_N
&
v\mathbf{A}_N+v\mathbf{B}_N
\\[4pt]
v\mathbf{C}_N+v\mathbf{D}_N
&
v\mathbf{C}_N+(1-v)\mathbf{D}_N
&
(1-v)\mathbf{C}_N+v\mathbf{D}_N
&
v\mathbf{C}_N+v\mathbf{D}_N
\end{bmatrix},
\end{align*}
where ${\bf I} =I_{2^{N-1}} $. 
Now, let 
\[
a_N = {\rm Tr} ({\bf A}_N ), \ b_N = {\rm Tr} ({\bf B}_N ), \ c_N = {\rm Tr} ({\bf C}_N ), \ d_N = {\rm Tr} ({\bf D}_N ) . 
\]
Then we have 
\begin{align*}
a_{N+1} & =  (1-v) a_N +v b_N +v c_N +v d_N , \\
b_{N+1} & =  v a_N +v b_N +v c_N +(1-v) d_N , \\
c_{N+1} & =  v a_N +v b_N +v c_N +(1-v) d_N , \\
d_{N+1} & =  (1-v) a_N +v b_N +v c_N +v d_N . 
\end{align*}
Note that 
\[
\mathbf{A}_N = \mathbf{D}_N
\quad \text{and} \quad
\mathbf{B}_N = \mathbf{C}_N,
\]
and so 
\[
Q^{(g)}_{N} = \left[ 
\begin{array}{cc}
{\bf A}_N & {\bf B}_N \\ 
{\bf B}_N & {\bf A}_N 
\end{array}  
\right] 
. 
\]
Then we get
\[
a_{N+1} = d_{N+1} , \ b_{N+1} = c_{N+1}.
\]
Thus, we obtain 
\begin{align*}
a_{N+1} & =  a_N +2v b_N , \\
b_{N+1} & =  a_N +2v b_N ,\\
a_2 &= b_2 =1
\end{align*}
and $a_{N+1} = b_{N+1} $.
So we have
\[
a_{N+1} =(1+2v) a_N, 
\]
and 
\[
a_{N+1} =(1+2v)^{N-1}  a_2 . 
\]
Then we obtain 
\[
a_{N+1} =(1+2v)^{N-1} . 
\]
Therefore, it follows that 
\[
{\rm Tr} (Q^{(g)}_{N} )= a_N + d_N =2 a_N =2 (1+2v)^{N-2} . 
\]
\end{proof}

\begin{rem}
For any square matrix $\mathbf Q$, we have
\[
\det(I-u\mathbf Q)=1-u\operatorname{Tr}(\mathbf Q)+O(u^2).
\]
Hence Theorem~4.1 yields
\[
\det\left(I_{2^N}-uQ_{N,f}^{(g)}\right)
=
1-2(1+2v)^{N-2}u+O(u^2).
\]
In particular, when $N=4$, we obtain
\[
\det\left(I_{16}-uQ_{4,f}^{(g)}\right)
=
1-2(1+2v)^2u+O(u^2).
\]
Thus the coefficient of $u$ in any explicit determinant
formula for $N=4$ must be $-2(1+2v)^2$.
This provides a useful consistency check for the
factorization in Theorem~8.1.
\end{rem}

\section{Zeta function for fuzzification of Rule 90} 

In this section, we derive a determinant factorization for the zeta function
of fuzzy Rule 90 by using a Hadamard block decomposition.  We first
decompose the configuration space according to the first coordinate.  With
respect to this decomposition, the global operator of fuzzy Rule 90 has
the symmetric block form
\[
Q_{N,f}^{(g)}
=
\left[ 
\begin{array}{cc} 
{\bf A}_N & {\bf B}_N\\
{\bf B}_N & {\bf A}_N
\end{array} 
\right] 
.
\]
Here \({\bf A}_N\) and \({\bf B}_N\) are \(2^{N-1}\times 2^{N-1}\) matrices.
The following theorem gives the corresponding determinant factorization.

\begin{thm}\label{thm:block-determinant}
Let $Q_{N,f}^{(g)}$
be the global operator of fuzzy Rule 90. Then, for \(N\geq2\),
\[
\det( I_{2^N}-uQ_{N,f}^{(g)})
=
\det( I_{2^{N-1}}-u(\mathbf A_N+\mathbf B_N))
\det( I_{2^{N-1}}-u(\mathbf A_N-\mathbf B_N)).
\]
Furthermore, 
\begin{align}
\begin{split}
& \det ( I_{2^{N-1}} -u( {\bf A}_N + {\bf B}_N )) \\
=&
\det ( I_{2^{N-2}} -u(2v+1)( {\bf A}_{N-1} + {\bf B}_{N-1} )) 
\det ( I_{2^{N-2}} -u(1-2v)( {\bf A}_{N-1} - {\bf B}_{N-1} )) .
\end{split}
\end{align}
\end{thm}

\begin{proof}
Define the Hadamard-type block matrix
\[
\mathbf H_N
=
\frac{1}{\sqrt{2}}
\left[ 
\begin{array}{cc} 
I_{2^{N-1}} & I_{2^{N-1}}\\
I_{2^{N-1}} & - I_{2^{N-1}}
\end{array} 
\right] 
.
\]
Then
\[
\mathbf H_N^{-1}=\mathbf H_N.
\]

A direct block matrix calculation gives
\[
\begin{aligned}
\mathbf H_N Q_{N,f}^{(g)}\mathbf  H_N
&=
\frac12
\left[ 
\begin{array}{cc} 
 I_{2^{N-1}} &  I_{2^{N-1}}\\
 I_{2^{N-1}} & - I_{2^{N-1}}
\end{array} 
\right] 
\left[ 
\begin{array}{cc} 
\mathbf A_N &\mathbf  B_N\\
\mathbf B_N & \mathbf A_N
\end{array} 
\right] 
\left[ 
\begin{array}{cc} 
I_{2^{N-1}} & I_{2^{N-1}}\\
 I_{2^{N-1}} & - I_{2^{N-1}} 
\end{array} 
\right] \\[2mm]
&=
\frac12
\left[ 
\begin{array}{cc} 
\mathbf A_N+\mathbf B_N & \mathbf A_N+\mathbf B_N\\
\mathbf A_N-\mathbf B_N &\mathbf  B_N-\mathbf A_N
\end{array} 
\right] 
\left[ 
\begin{array}{cc} 
 I_{2^{N-1}} & I_{2^{N-1}}\\
 I_{2^{N-1}} & -I_{2^{N-1}} 
\end{array} 
\right] \\[2mm]
&=
\frac12
\left[ 
\begin{array}{cc} 
2(\mathbf A_N+\mathbf B_N) & 0\\
0 & 2(\mathbf A_N-\mathbf B_N)
\end{array} 
\right] 
.
\end{aligned}
\]
Therefore,
\[
\mathbf H_N Q_{N,f}^{(g)}\mathbf  H_N
=
\left[ 
\begin{array}{cc} 
\mathbf A_N+\mathbf B_N & 0\\
0 & \mathbf A_N-\mathbf B_N
\end{array} 
\right] 
.
\]
Hence,
\[
\begin{aligned}
\mathbf H_N\left( I_{2^N}-uQ_{N,f}^{(g)}\right)\mathbf H_N
&=
I_{2^N}
-
u\mathbf H_N Q_{N,f}^{(g)}\mathbf H_N \\[1mm]
&=
\left[ 
\begin{array}{cc} 
 I_{2^{N-1}}-u(\mathbf A_N+\mathbf B_N) & 0\\
0 &  I_{2^{N-1}}-u(\mathbf A_N-\mathbf B_N)
\end{array} 
\right] 
.
\end{aligned}
\]

Since \(\mathbf H_N^{-1}=\mathbf H_N\), this is a similarity transformation. Therefore, taking determinants gives
\[
\det\left( I_{2^N}-uQ_{N,f}^{(g)}\right)
=
\det\left( I_{2^{N-1}}-u(\mathbf A_N+\mathbf B_N)\right)
\det\left(I_{2^{N-1}}-u(\mathbf A_N-\mathbf B_N)\right).
\]

By the definition of the IPS-type zeta function,
\[
\zeta(Q_f^{(l)},\PM_N,u)
=
\det\left(I_{2^N}-uQ_{N,f}^{(g)}\right)^{-1/2^N}.
\]
Thus,
\[
\zeta(Q_f^{(l)},\PM_N,u)^{-1}
=
\det\left( I_{2^N}-uQ_{N,f}^{(g)}\right)^{1/2^N}.
\]
Substituting the determinant factorization above, we obtain
\[
\zeta(Q_f^{(l)},\PM_N,u)^{-1}
=
\left[
\det\left(I_{2^{N-1}}-u(\mathbf A_N+\mathbf B_N)\right)
\det\left(I_{2^{N-1}}-u(\mathbf A_N-\mathbf B_N)\right)
\right]^{1/2^N}.
\] 
In the proof of Theorem 4.1, we have
\begingroup
\setlength{\arraycolsep}{4pt}
\renewcommand{\arraystretch}{1.3}
\begin{align}
&\begin{bmatrix}
\mathbf{A}_N & \mathbf{B}_N \\
\mathbf{C}_N & \mathbf{D}_N
\end{bmatrix}
=
\begin{bmatrix}
\mathbf{A}_{N} & \mathbf{B}_{N} \\
\mathbf{B}_{N} & \mathbf{A}_{N}
\end{bmatrix}
\notag\\[10pt]
&\resizebox{0.98\linewidth}{!}{$\displaystyle
{}=
\begin{bmatrix}
(1-v)\mathbf{A}_{N-1}+v\mathbf{B}_{N-1}
&
v\mathbf{A}_{N-1}+v\mathbf{B}_{N-1}
&
v\mathbf{A}_{N-1}+v\mathbf{B}_{N-1}
&
v\mathbf{A}_{N-1}+(1-v)\mathbf{B}_{N-1}
\\[4pt]
v\mathbf{A}_{N-1}+(1-v)\mathbf{B}_{N-1}
&
v\mathbf{A}_{N-1}+v\mathbf{B}_{N-1}
&
v\mathbf{A}_{N-1}+v\mathbf{B}_{N-1}
&
(1-v)\mathbf{A}_{N-1}+v\mathbf{B}_{N-1}
\\[4pt]
v\mathbf{A}_{N-1}+v\mathbf{B}_{N-1}
&
v\mathbf{A}_{N-1}+(1-v)\mathbf{B}_{N-1}
&
(1-v)\mathbf{A}_{N-1}+v\mathbf{B}_{N-1}
&
v\mathbf{A}_{N-1}+v\mathbf{B}_{N-1}
\\[4pt]
v\mathbf{A}_{N-1}+v\mathbf{B}_{N-1}
&
(1-v)\mathbf{A}_{N-1}+v\mathbf{B}_{N-1}
&
v\mathbf{A}_{N-1}+(1-v)\mathbf{B}_{N-1}
&
v\mathbf{A}_{N-1}+v\mathbf{B}_{N-1}
\end{bmatrix}.
$}
\end{align}
\endgroup
Therefore we get   
\[
{\bf A}_N + {\bf B}_N 
= \left[ 
\begin{array}{cc}
{\bf A}_{N-1} +2v {\bf B}_{N-1} & 2v {\bf A}_{N-1} + {\bf B}_{N-1} \\
2v {\bf A}_{N-1} + {\bf B}_{N-1} & {\bf A}_{N-1} +2v {\bf B}_{N-1} 
\end{array}  
\right] 
. 
\] 
Thus, we obtain  
\begin{align*}
& {\bf H} ( {\bf A}_N + {\bf B}_N ) {\bf H} \\
=
& 
\left[ 
\begin{array}{cc}
(1+2v)( {\bf A}_{N-1} + {\bf B}_{N-1} ) & {\bf 0} \\
{\bf 0} & (1-2v)( {\bf A}_{N-1} - {\bf B}_{N-1} )
\end{array}  
\right] 
.
\end{align*}
Therefore, it follows that 
\begin{align}
\begin{split}
& \det ( I_{2^{N-1}} -u( {\bf A}_N + {\bf B}_N )) \\
=&
\det ( I_{2^{N-2}} -u(2v+1)( {\bf A}_{N-1} + {\bf B}_{N-1} )) 
\det ( I_{2^{N-2}} -u(1-2v)( {\bf A}_{N-1} - {\bf B}_{N-1} )) .
\end{split}
\end{align}
This completes the proof.
\end{proof}

By Theorem 5.1, we obtain a recurrence formula for $\det( I_{2^N}-uQ_{N,f}^{(g)})$.

\begin{cor} 
\begin{align*}
& \overline{\zeta } (Q^{(l)}_f , \PM_N ,u) \\
=
&  \left\{ 1-u^2(2v+1)^{2(N-3)} (1-2v)^4 \right\}^{-1/2^{N}} 
\left\{ 1+u(2v+1)^{N-2} (2v-1)\right\}^{-1/2^N} 
\left\{1-u(2v+1)^{N-1} \right\}^{-1/2^N} \\
& \times  \prod^{N-1}_{j=3} \left\{ \det (I_{2^{j-1}} -u(2v+1)^{N-j-1} (1-2v)( {\bf A}_j - {\bf B}_j )\right\}^{-1/2^N} 
\times \left\{ \det ( I_{2^{N-1}} -u( {\bf A}_N - {\bf B}_N ))\right\}^{-1/2^N} .  
\end{align*}
\end{cor} 

\begin{proof} By 
\[
{\bf A}_{N-1} + {\bf B}_{N-1} 
= \left[ 
\begin{array}{cc}
{\bf A}_{N-2} +2v {\bf B}_{N-2} & 2v {\bf A}_{N-2} + {\bf B}_{N-2} \\
2v {\bf A}_{N-2} + {\bf B}_{N-2} & {\bf A}_{N-2} +2v {\bf B}_{N-2} 
\end{array}  
\right] 
,
\]
and (7), we have 
\begin{align*}
& \det ( I_{2^{N-2}} -u(1+2v)( {\bf A}_{N-1} + {\bf B}_{N-1} )) \\
=&
\det ( I_{2^{N-3}} -u(2v+1)^2 ( {\bf A}_{N-2} + {\bf B}_{N-2} )) 
\det ( I_{2^{N-3}} -u(1-2v)(1+2v)( {\bf A}_{N-2} - {\bf B}_{N-2} )) .
\end{align*}
From Theorem 5.1, we obtain 
\begin{align*}
& \det ( I_{2^N} -u Q^{(g)}_{N} ) \\
=&
\det ( I_{2^{N-3}} -u(2v+1)^2 ( {\bf A}_{N-2} + {\bf B}_{N-2} )) 
\det ( I_{2^{N-3}} -u(1-2v)(1+2v)( {\bf A}_{N-2} - {\bf B}_{N-2} )) \\
& \times
\det ( I_{2^{N-2}} -u (1-2v)({\bf A}_{N-1} - {\bf B}_{N-1} )) 
\det ( I_{2^{N-1}} -u ({\bf A}_N - {\bf B}_N )) \\
=&
\det ( I_{2^{N-4}} -u(2v+1)^3 ( {\bf A}_{N-3} + {\bf B}_{N-3} )) 
\det ( I_{2^{N-4}} -u(1-2v)(1+2v)^2 ( {\bf A}_{N-3} - {\bf B}_{N-3} )) \\
& \times
\det ( I_{2^{N-3}} -u(2v+1)(1-2v) ( {\bf A}_{N-2} - {\bf B}_{N-2} )) 
\det ( I_{2^{N-2}} -u(1-2v)( {\bf A}_{N-1} - {\bf B}_{N-1} )) \\
& \times
\det ( I_{2^{N-1}} -u ({\bf A}_N - {\bf B}_N )) . 
\end{align*}
Inductively, we have 
\begin{align*}
& \det ( I_{2^N} -u Q^{(g)}_{N} ) \\
=&
\det ( I_{2} -u(2v+1)^{N-2} ( {\bf A}_2 + {\bf B}_2 )) 
\det ( I_{2} -u(1-2v)(1+2v)^{N-3} ( {\bf A}_2 - {\bf B}_2 )) \\
& \times
\det ( I_{2^2} -u(1-2v)(1+2v)^{N-4}  ({\bf A}_3 - {\bf B}_3 ))  \cdots \\
& \times
\det ( I_{2^{N-3}} -u(2v+1)(1-2v) ( {\bf A}_{N-2} - {\bf B}_{N-2} )) 
\det ( I_{2^{N-2}} -u(1-2v)( {\bf A}_{N-1} - {\bf B}_{N-1} )) \\
& \times
\det ( I_{2^{N-1}} -u({\bf A}_N - {\bf B}_N )) . 
\end{align*}
Since 
\[
{\bf A}_2 + {\bf B}_2 
= \left[ 
\begin{array}{cc}
1 & 2v \\
2v & 1  
\end{array}  
\right]
,\quad
{\bf A}_2 - {\bf B}_2 
= \left[ 
\begin{array}{cc}
1-2v & 0 \\
0 & 2v-1 
\end{array}  
\right] 
, 
\]
we have 
\[
\det ( I_{2} -u(2v+1)^{N-2} ( {\bf A}_2 + {\bf B}_2 ))=(1+u(2v+1)^{N-2} (2v-1))(1-u(2v+1)^{N-1} ) 
\]
and 
\[
\det ( ( I_{2} -u(1-2v)(1+2v)^{N-3} ( {\bf A}_2 - {\bf B}_2 ))=(1-u^2(2v+1)^{2(N-3)} (1-2v)^4 ) . 
\]
Hence, the result follows. 
\end{proof}

\section{Calculation for $\det ( I_{2^{N-1}} -u({\bf A}_N - {\bf B}_N ))$} 

By Theorem 5.1, the determinant
\(\det(I_{2^N}-uQ_{N,f}^{(g)})\) is reduced to the study of
\(\det(I_{2^{N-1}}-u(\mathbf A_N+\mathbf B_N))\) and \(\det(I_{2^{N-1}}-u(\mathbf A_N -\mathbf B_N))\).
In this section, we derive a recursive formula for the latter determinant.
Put
\[
\mathbf S_N=\mathbf A_N+\mathbf B_N,
\qquad
\mathbf D_N=\mathbf A_N-\mathbf B_N,
\qquad
w=1-2v.
\]
The following result holds for $\det ( I_{2^{N-1}} -u({\bf \mathbf A}_N - {\bf \mathbf B}_N ))$. 
\begin{thm}\label{thm:DN-recursion}
For \(N\geq3\), we have
\[
\det(I_{2^{N-1}}-u\mathbf D_N)
=
\det(I_{2^{N-2}}-u^2w^2\mathbf D_{N-1}\mathbf S_{N-1}).
\]
Equivalently,
\[
\det(I_{2^{N-1}}-u(\mathbf A_N-\mathbf B_N))
=
\det\left(
I_{2^{N-2}}
-u^2(1-2v)^2
(\mathbf A_{N-1}-\mathbf B_{N-1})(\mathbf A_{N-1}+\mathbf B_{N-1})
\right).
\]
\end{thm}

\begin{proof}
Indeed, (8) implies that the blocks \(\mathbf A_N\) and \(\mathbf B_N\) satisfy
\[
\mathbf A_N=
\left[ 
\begin{array}{cc} 
(1-v)\mathbf A_{N-1}+v\mathbf B_{N-1} & v\mathbf A_{N-1}+v\mathbf B_{N-1}\\
v\mathbf A_{N-1}+(1-v)\mathbf B_{N-1} & v\mathbf A_{N-1}+v\mathbf B_{N-1}
\end{array} 
\right] 
,
\]
and
\[
\mathbf B_N= 
\left[ 
\begin{array}{cc} 
v\mathbf A_{N-1}+v\mathbf B_{N-1} & v\mathbf A_{N-1}+(1-v)\mathbf B_{N-1}\\
v\mathbf A_{N-1}+v\mathbf B_{N-1} & (1-v)\mathbf A_{N-1}+v\mathbf B_{N-1}
\end{array} 
\right] 
.
\]
From the block recurrence of \(\mathbf A_N\) and \(\mathbf B_N\), we have
\[
\mathbf D_N=\mathbf A_N-\mathbf B_N
=
w
\left[ 
\begin{array}{cc} 
\mathbf A_{N-1} & -\mathbf B_{N-1}\\
\mathbf B_{N-1} & -\mathbf A_{N-1}
\end{array} 
\right] ,
\]
where $w=1-2v$.

Let
\[
\mathbf H=
\frac1{\sqrt2}
\left[ 
\begin{array}{cc} 
I_{2^{N-2}} & I_{2^{N-2}}\\
I_{2^{N-2}} & -I_{2^{N-2}}
\end{array} 
\right] 
.
\]
Then \(\mathbf H^{-1}=\mathbf H\), and a direct calculation gives
\[
\mathbf H\mathbf D_N\mathbf H
=
w
\left[ 
\begin{array}{cc} 
0 & \mathbf A_{N-1}+\mathbf B_{N-1}\\
\mathbf A_{N-1}-\mathbf B_{N-1} & 0
\end{array} 
\right] 
=
w
\left[ 
\begin{array}{cc} 
0 & \mathbf S_{N-1}\\
\mathbf D_{N-1} & 0
\end{array} 
\right] 
.
\]
Therefore,
\[
\begin{aligned}
\det(I_{2^{N-1}}-u\mathbf D_N)
&=
\det
\left[ 
\begin{array}{cc} 
I_{2^{N-2}} & -uw\mathbf S_{N-1}\\
-uw\mathbf D_{N-1} & I_{2^{N-2}}
\end{array} 
\right] 
.
\end{aligned}
\]
Using
\[
\det
\left[ 
\begin{array}{cc} 
 I & -\mathbf X\\
-\mathbf Y &  I
\end{array} 
\right] 
=
\det(I-\mathbf Y\mathbf X),
\]
we obtain
\[
\det(I_{2^{N-1}}-u\mathbf D_N)
=
\det(I_{2^{N-2}}-u^2w^2\mathbf D_{N-1}\mathbf S_{N-1}).
\]
This completes the proof.
\end{proof}

\begin{rem}
The determinant recursion in Theorem~\ref{thm:DN-recursion} shows that the
Hadamard block decomposition can be iterated.  Therefore, for fuzzy Rule
90, the zeta function admits a recursive determinant description.

However, when \(v\neq 0\), the global operator \(Q_{N,f}^{(g)}\) is no longer
a permutation matrix.  Hence the fixed-point counting argument used for the
ordinary Rule 90 cannot be applied directly.  In particular, the dyadic product
formula obtained in Theorem~\ref{thm:Rule90-dyadic} should be regarded as a
special feature of the deterministic case \(v=0\) \ (Rule 90).

Thus, for general \(0<v<1\), the natural analogue of the dyadic formula is not
a closed dyadic product formula, but a recursive determinant formula.
\end{rem}

\section{Zeta function for fuzzification of Rule 90 in the case of $N=3$} 

The following result holds for $\overline{\zeta } (Q^{(l)}_f , \PM_3 ,u)$.

\begin{thm}
\begin{align*}
& \overline{\zeta } (Q^{(l)}_f , \PM_3 ,u) \\
=& 
(1-u(1-4 v^2 ))^{-1/8} (1-u(2v+1)^2 )^{-1/8} (1-u(1-2v)^2 )^{-1/8} (1+u(1-2v)^2 )^{-1/8} \\
& \times
(1- u^4 (1-2v)^7 (1+2v))^{-1/8} . 
\end{align*}
\end{thm} 

\begin{proof}  By Theorem 5.1, we have 
\begin{align*}
\det ( I_8 -u Q^{(g)}_{3} ) =&
\det ( I_{2} -u(2v+1)( {\bf A}_2 + {\bf B}_2 )) 
\det ( I_{2} -u(1-2v)( {\bf A}_2 - {\bf B}_2 )) \\
&  \times  \det ( I_4 -u({\bf A}_3 - {\bf B}_3 )) . 
\end{align*}
Since 
\[
{\bf A}_2 = \left[ 
\begin{array}{cc}
1-v & v \\
v & v 
\end{array}  
\right] 
, \quad
{\bf B}_2 = \left[ 
\begin{array}{cc}
v & v \\
v & 1-v 
\end{array}  
\right] 
, 
\]
we have 
\begin{align*}
\det ( I_{2} -u(2v+1)( {\bf A}_2 + {\bf B}_2 ))
&=
\det 
\left[ 
\begin{array}{cc}
1-u(2v+1) & -2uv(2v+1) \\
-2uv(2v+1) & 1-u(2v+1) 
\end{array}  
\right] \\
&= (1-u(2v+1))^2 -4u^2 v^2 (2v+1)^2 \\ 
&= (1-u(2v+1)(1-2v))(1-u(2v+1)^2 ) . 
\end{align*}
Furthermore, we have 
\begin{align*}
\det ( I_{2} -u(1-2v)( {\bf A}_2 - {\bf B}_2 )) &=
\det 
\left[ 
\begin{array}{cc}
1-u(1-2v)^2 & 0 \\
0 & 1+u(1-2v)^2 
\end{array}  
\right] \\
&=  (1-u(1-2v)^2)(1+u(1-2v)^2) . 
\end{align*}

Now, we have 
\[
{\bf A}_3 - {\bf B}_3 = 
\left[ 
\begin{array}{cc}
(1-2v) {\bf A}_2 & -(1-2v) {\bf B}_2 \\
(1-2v) {\bf B}_2 & -(1-2v) {\bf A}_2 
\end{array}  
\right] 
. 
\]
By Theorem 6.1, we obtain
\begin{align*}
\det ( I_4 -u({\bf A}_3 - {\bf B}_3 )) &=
\det ( I_2 -u^2 (1-2v)^2 ({\bf A}_2 - {\bf B}_2 )({\bf A}_2 + {\bf B}_2 )) . 
\end{align*}
Since 
\begin{align*}
({\bf A}_2 - {\bf B}_2 )({\bf A}_2 + {\bf B}_2 )
=(1-2v) 
\left[ 
\begin{array}{cc}
1 & 2v \\ 
-2v & -1 
\end{array}  
\right] 
, 
\end{align*}
we have 
\begin{align*}
\det ( I_4 -u({\bf A}_3 - {\bf B}_3 ))
&=
\det \left[ 
\begin{array}{cc}
1-u^2(1-2v)^3 & -2vu^2 (1-2v)^3 \\
2vu^2 (1-2v)^3 & 1+u^2(1-2v)^3 
\end{array}  
\right] \\
&= (1-u^2(1-2v)^3 )(1+u^2(1-2v)^3 )+4v^2 u^4 (1-2v)^6 \\ 
&= 1-u^4 (1-2v)^6 + 4v^2 u^4 (1-2v)^6 \\
&= 1- u^4(1 - 2v)^6 (1 - 4v^2) \\
&= 1- u^4(1 - 2v)^7 (1 + 2v) .
\end{align*}
Therefore, it follows that 
\begin{align*}
\overline{\zeta } (Q^{(l)}_f , \PM_3 ,u) =&
(1-u(1-4 v^2 ))^{-1/8} (1-u(2v+1)^2 )^{-1/8} (1-u(1-2v)^2 )^{-1/8} 
(1+u(1-2v)^2 )^{-1/8} \\
& \times 
(1- u^4 (1-2v)^7 (1+2v))^{-1/8} . 
\end{align*}

\end{proof}

\section{Determinant Calculations Using Hadamard Block Decomposition}

In this section, we give further calculations of $\det ( {\bf I}_{2^{N-1}} -u({\bf A}_N - {\bf B}_N ))$ by the Hadamard block matrix.

\begin{pro}
\begin{align*}
&\det\bigl(
I_{2^{N-1}}-u(\mathbf{A}_N-\mathbf{B}_N)
\bigr)
\\[4pt]
&=\det\Bigl(
I_{2^{N-3}}
-u^4(1-2v)^7(1+2v)
(\mathbf{A}_{N-2}+\mathbf{B}_{N-2})
(\mathbf{A}_{N-2}-\mathbf{B}_{N-2})^2
(\mathbf{A}_{N-2}+\mathbf{B}_{N-2})
\Bigr).
\end{align*}
\end{pro}

\begin{proof}
Let \( w= 1 - 2v \) and
\[
\begin{aligned}
\mathbf{P}_{N-1}
&=(\mathbf{A}_{N-1}-\mathbf{B}_{N-1})
  (\mathbf{A}_{N-1}+\mathbf{B}_{N-1})
\\[4pt]
&=\mathbf{D}_{N-1}\mathbf{S}_{N-1}.
\end{aligned}
\]
Then we can express
\[
\mathbf P_{N-1} = 
w
\begin{bmatrix}
\mathbf S_{N-2} & \mathbf T_{N-2} \\
-\mathbf T_{N-2} & -\mathbf S_{N-2}
\end{bmatrix} 
,\quad \ w=1-2v ,
\]
where
\[
\mathbf S_{N-2} = \mathbf A^2_{N-2} -\mathbf B^2_{N-2} + 2v(\mathbf A_{N-2} \mathbf B_{N-2} - \mathbf B_{N-2} \mathbf A_{N-2}),
\]
\[
\mathbf T_{N-2} = 2v(\mathbf A^2_{N-2} - \mathbf B^2_{N-2} )+ (\mathbf A_{N-2}\mathbf B_{N-2} - \mathbf B_{N-2} \mathbf A_{N-2}).
\]

\bigskip

\noindent
Hence, by Theorem 6.1, we have 
\[
\begin{aligned}
\det\!\bigl(I_{2^{N-1}}-u\mathbf{D}_N\bigr)
&=\det\!\left(
I_{2^{N-2}}
-u^2w^2\,\mathbf{D}_{N-1}\mathbf{S}_{N-1}
\right).
\end{aligned}
\]

\bigskip
Set
\[
\alpha = u^{2} w^{3}, \qquad
\mathbf{M} =
\begin{bmatrix}
\mathbf{S}_{N-2} & \mathbf{T}_{N-2} \\
-\mathbf{T}_{N-2} & -\mathbf{S}_{N-2}
\end{bmatrix}
= \frac{\mathbf {P_{N-1}}}{w}.
\]

\noindent
Then we have 
\[
\mathbf H\mathbf M \mathbf H = \begin{bmatrix}
    0 & \mathbf S_{N-2} - \mathbf T_{N-2} \\
\mathbf S_{N-2} +\mathbf T_{N-2} & 0
\end{bmatrix}.
\]
Thus we get
\[
\begin{aligned}
\det\!\left(I_{2^{N-2}} - \alpha \mathbf M\right)
&= \det
\begin{bmatrix}
I_{2^{N-3}} & -\alpha\!\left(\mathbf S_{N-2}-\mathbf T_{N-2}\right) \\
-\alpha\!\left(\mathbf S_{N-2}+\mathbf T_{N-2}\right) & I_{2^{N-3}}
\end{bmatrix} \\
&= \det\!\left(
I_{2^{N-3}} - \alpha^{2}\,(\mathbf S_{N-2}-\mathbf T_{N-2})(\mathbf S_{N-2}+ \mathbf T_{N-2})
\right).
\end{aligned}
\]

\noindent
By expanding $\mathbf S_{N-2}$ and $\mathbf T_{N-2}$, we have
\[
\begin{aligned}
\mathbf S_{N-2} -\mathbf T_{N-2}
&= (\mathbf A_{N-2}^2 -\mathbf B_{N-2}^2)
+ 2v(\mathbf A_{N-2} \mathbf B_{N-2} - \mathbf B_{N-2} \mathbf A_{N-2}) \\
&\quad - \Bigl[
(\mathbf A_{N-2}\mathbf B_{N-2} - \mathbf B_{N-2} \mathbf A_{N-2})
+ 2v(\mathbf A_{N-2}^2 -\mathbf B_{N-2}^2)
\Bigr] \\
&= (1 - 2v)\bigl[
(\mathbf A_{N-2}^2 - \mathbf B_{N-2}^2)
- (\mathbf A_{N-2} \mathbf B_{N-2} - \mathbf B_{N-2} \mathbf A_{N-2})
\bigr].
\end{aligned}
\]

\[
\mathbf S_{N-2} +\mathbf T_{N-2}
= (1+2v)\bigl[(\mathbf A_{N-2}^2 - \mathbf B_{N-2}^2)
+ (\mathbf A_{N-2} \mathbf B_{N-2} - \mathbf B_{N-2} \mathbf A_{N-2})\bigl].
\]

\bigskip

\noindent
We recall the standard matrix identities:
\[
\begin{aligned}
(\mathbf{A}+\mathbf{B})(\mathbf{A}-\mathbf{B})
&=\mathbf{A}^2-\mathbf{B}^2
-(\mathbf{A}\mathbf{B}-\mathbf{B}\mathbf{A}),
\\[4pt]
(\mathbf{A}-\mathbf{B})(\mathbf{A}+\mathbf{B})
&=\mathbf{A}^2-\mathbf{B}^2
+(\mathbf{A}\mathbf{B}-\mathbf{B}\mathbf{A}).
\end{aligned}
\]

\noindent
Hence, we obtain the compact forms:
\[
\mathbf S_{N-2} -\mathbf T_{N-2} = (1-2v) (\mathbf A_{N-2} +\mathbf B_{N-2})(\mathbf A_{N-2} -\mathbf B_{N-2}), 
\]
\[
\mathbf S_{N-2} +\mathbf T_{N-2} = (1+2v) (\mathbf A_{N-2} - \mathbf B_{N-2})(\mathbf A_{N-2} +\mathbf B_{N-2}).
\]

\bigskip

\noindent
Substituting these into the determinant expression gives
\begin{align*}
&\det\bigl(
I_{2^{N-1}}-u(\mathbf{A}_N-\mathbf{B}_N)
\bigr)
\\[4pt]
&=\det\Bigl(
I_{2^{N-3}}
-u^4w^7(1+2v)
(\mathbf{A}_{N-2}+\mathbf{B}_{N-2})
(\mathbf{A}_{N-2}-\mathbf{B}_{N-2})^2
(\mathbf{A}_{N-2}+\mathbf{B}_{N-2})
\Bigr).
\end{align*}
\noindent
Therefore, the recursive determinant identity is established. 
\end{proof}

\begin{pro} 
\begin{align*}
& \det \left( {I}_{2^{N-1}} - u ({\bf A}_N - {\bf B}_N) \right) \\
&= \det \Bigl( I_{2^{N-4}} - u^4 (1 - 2v)^9 (1 + 2v)^3 (\mathbf A_{N-3} + \mathbf B_{N-3})(\mathbf A_{N-3} 
- \mathbf B_{N-3})(\mathbf A_{N-3} + \mathbf B_{N-3})^2 \Bigr) \\
& \times \det \Bigl( I_{2^{N-4}} - u^4 (1-2v)^{11} (1 + 2v) (\mathbf A_{N-3} - \mathbf B_{N-3})(\mathbf A_{N-3} 
+ \mathbf B_{N-3})(\mathbf A_{N-3} - \mathbf B_{N-3})^2 \Bigr).
\end{align*}

\end{pro} 

\begin{proof}
We consider the recursive determinant:
\[
\det\left( I_{2^{N-1}} - u (\mathbf A_N -\mathbf B_N) \right)
= \det\left( I_{2^{N-3}} - \beta \mathbf M_{N-2} \right),
\]
where
\[
\beta := u^4 w^7 (1 + 2v),
\ w = 1 - 2v.
\]
and
\[
\mathbf M_{N-2}
= (\mathbf A_{N-2} +\mathbf B_{N-2})(\mathbf A_{N-2} -\mathbf B_{N-2})^2(\mathbf A_{N-2} +\mathbf B_{N-2}).
\]

\bigskip

\noindent
From the proof of Proposition 8.1, we know
\[
\mathbf A_{k} +\mathbf  B_{k} =
\begin{bmatrix}
\mathbf A_{k-1} + 2v\mathbf B_{k-1} & 2v\mathbf A_{k-1} +\mathbf B_{k-1} \\
2v\mathbf A_{k-1} +\mathbf B_{k-1} & \mathbf A_{k-1} + 2v \mathbf B_{k-1}
\end{bmatrix},
\]
and
\[
\mathbf A_{k} - \mathbf B_{k} = w
\begin{bmatrix}
\mathbf A_{k-1} & -\mathbf B_{k-1} \\
\mathbf B_{k-1} & -\mathbf A_{k-1}
\end{bmatrix} .
\]

\bigskip

\noindent
Let
\[
\mathbf X :=\mathbf A_{N-3}, \qquad \mathbf Y := \mathbf B_{N-3}.
\]

Then we define
\[
\mathbf A_{N-2} +\mathbf B_{N-2} = 
\begin{bmatrix}
\mathbf P &\mathbf Q \\
\mathbf Q &\mathbf P
\end{bmatrix},
\qquad
\mathbf A_{N-2} -\mathbf B_{N-2} =
w
\begin{bmatrix}
\mathbf X & -\mathbf Y \\
\mathbf Y & -\mathbf X
\end{bmatrix},
\]
where
\[
\mathbf P := \mathbf X + 2v\mathbf Y, \qquad \mathbf Q := 2v\mathbf X + \mathbf Y.
\]

\bigskip

\noindent

We introduce the Hadamard block matrix
\[
\mathbf H = \frac{1}{\sqrt{2}}
\begin{bmatrix}
I & I \\
I & -I
\end{bmatrix},
\quad I = I_{2^{N-4}}.
\]
For the symmetric part of \(\mathbf A_{N-2} + \mathbf B_{N-2} \), we have 
\[
\mathbf H (\mathbf A_{N-2} +\mathbf B_{N-2})\mathbf H =
\begin{bmatrix}
\mathbf P +\mathbf Q & 0 \\
0 & \mathbf P -\mathbf Q
\end{bmatrix}.
\]
For the symmetric part of \(\mathbf A_{N-2} -\mathbf B_{N-2} \), we have 
\[
\mathbf H (\mathbf A_{N-2} -\mathbf B_{N-2})\mathbf H =w
\begin{bmatrix}
0 &\mathbf X +\mathbf Y \\
\mathbf X -\mathbf Y & 0
\end{bmatrix}.
\]
Thus we get
\[
\mathbf H (\mathbf A_{N-2} -\mathbf B_{N-2})^2 \mathbf H 
= w^2 
\begin{bmatrix}
(\mathbf X +\mathbf Y)(\mathbf X -\mathbf Y) & 0 \\
0 & (\mathbf X -\mathbf Y)(\mathbf X +\mathbf Y)
\end{bmatrix}.
\]

Then,
\begin{align*}
\mathbf H \mathbf M_{N-2} \mathbf H 
&= 
\begin{bmatrix}
\mathbf P +\mathbf Q & 0 \\
0 & \mathbf P -\mathbf Q
\end{bmatrix}
w^2
\begin{bmatrix}
(\mathbf X +\mathbf Y)(\mathbf X -\mathbf Y) & 0 \\
0 & (\mathbf X -\mathbf Y)(\mathbf X +\mathbf Y)
\end{bmatrix}
\begin{bmatrix}
\mathbf P +\mathbf Q & 0 \\
0 & \mathbf P -\mathbf Q
\end{bmatrix} \\[6pt]
&= 
w^2
\begin{bmatrix}
(\mathbf P +\mathbf Q)(\mathbf X +\mathbf Y)(\mathbf X -\mathbf Y)(\mathbf P +\mathbf Q) & 0 \\
0 & (\mathbf P -\mathbf Q)(\mathbf X -\mathbf Y)(\mathbf X +\mathbf Y)(\mathbf P -\mathbf Q)
\end{bmatrix}.
\end{align*}
Thus, we get 
\[
\begin{aligned}
& \det\!\left( I_{2^{N-3}} - \beta w^{2} \mathbf M_{N-2} \right)
= \det\!\left( I_{2^{N-3}} - \beta \omega^{2} \mathbf H\mathbf M_{N-2} \mathbf H \right) \\[4pt]
&= \det\!\left(
I_{2^{N-3}}
- \beta w^2
\begin{bmatrix}
(\mathbf P +\mathbf Q)(\mathbf X +\mathbf Y)(\mathbf X -\mathbf Y)(\mathbf P +\mathbf Q) & 0 \\[3pt]
0 & (\mathbf P -\mathbf Q)(\mathbf X -\mathbf Y)(\mathbf X +\mathbf Y)(\mathbf P -\mathbf Q)
\end{bmatrix}
\right) \\[6pt]
&= 
\det\!\left( I_{2^{N-4}} - \beta w^2 (\mathbf P +\mathbf Q)(\mathbf X +\mathbf Y)(\mathbf X -\mathbf Y)(\mathbf P +\mathbf Q) \right) \\ 
& \times \det\!\left( I_{2^{N-4}} - \beta w^2 (\mathbf P -\mathbf Q)(\mathbf X -\mathbf Y)(\mathbf X +\mathbf Y)(\mathbf P -\mathbf Q) \right).
\end{aligned}
\]

Since
\[
\mathbf P +\mathbf Q = (1 + 2v)(\mathbf A_{N-3} + \mathbf B_{N-3}), \quad
\mathbf P -\mathbf Q = (1 - 2v)(\mathbf A_{N-3} - \mathbf B_{N-3}),
\]
we obtain
\begin{align*}
& \det \left( I_{2^{N-4}} - \beta w^2(\mathbf P +\mathbf Q)(\mathbf X +\mathbf Y)(\mathbf X -\mathbf Y)(\mathbf P +\mathbf Q) \right) \\
&= \det \left( I_{2^{N-4}} - u^4 (1 - 2v)^9 (1 + 2v)^3 
(\mathbf A_{N-3} +\mathbf B_{N-3})^2(\mathbf A_{N-3} -\mathbf B_{N-3})(\mathbf A_{N-3} +\mathbf B_{N-3}) \right).
\end{align*}

We also have
\begin{align*}
& \det \left( I_{2^{N-4}} - \beta w^2 (\mathbf P -\mathbf Q)(\mathbf X -\mathbf Y)(\mathbf X +\mathbf Y)(\mathbf P -\mathbf Q) \right) \\
&= \det \left( I_{2^{N-4}} - u^4 (1 + 2v)(1 - 2v)^{11} 
(\mathbf A_{N-3} - \mathbf B_{N-3})^2(\mathbf A_{N-3} + \mathbf B_{N-3})(\mathbf A_{N-3} - \mathbf B_{N-3}) \right).
\end{align*}
\end{proof}

Now, we present a formula of $\det\!\left(I_{2^{4}} - u\,Q_{4}^{(g)}\right)$.

\begin{thm}[Case of $N=4$]
\[
\begin{aligned}
\det\!\left(I_{2^{4}} - u\,Q_{4}^{(g)}\right)
&= \bigl(1 - u^2 (1 + 2v)^2 (1 - 2v)^{4}\bigr) 
   \bigl(1 - u(1 + 2v)^{2}(1 - 2v)\bigr)  \bigl(1 - u(1 + 2v)^3\bigr) \\
&\quad \times \bigl(1 - u^4 (1 - 2v)^{11} (1+2v) \bigr)^2 \bigl(1 - u^{4}(1 - 2v)^{9}(1 + 2v)^{3}\bigr).
\end{aligned}
\]
\end{thm}

\begin{proof}
By the definition of $\overline{\zeta}$,
\[
\det\!\left(I_{2^{4}} - u\,Q_{4}^{(g)}\right)
= \overline{\zeta}\!\left(Q_{4}^{(l)},\,\PM_{4},\,u\right)^{-16}.
\]
From Corollary 5.1, we have
\begin{align*}
\det\!\left(I_{2^{4}} - u\,Q_{4}^{(g)}\right)
&= \bigl(1 - u^2 (1+2v)^2 (1-2v)^{4}\bigr) 
   \bigl(1 + u(1+2v\bigr)^{2}
   (2v-1)\bigr) 
   \bigl(1 - u(2v+1)^3\bigr) \\
&\quad \times
\det\!\left(I_{2^{2}} - u(2v+1)^{0}(1-2v)(\mathbf A_{3}- \mathbf B_{3})\right)
\det\!\left(I_{2^{3}} - u(\mathbf A_{4}- \mathbf B_{4})\right).
\end{align*}
By Proposition 8.1 and Proposition 8.2, we have
\begin{align*}
\det\!\left(I_{2^{4}} - u\,Q_{4}^{(g)}\right)
&= \bigl(1 - u^2 (1+2v)^2 (1-2v)^{4}\bigr) 
   \bigl(1 + u(1+2v\bigr)^{2}
   (2v-1)\bigr)
   \bigl(1 - u(2v+1)^3\bigr) \\
&\quad \times
\det\!\left(I_{2^0} - u^{4}(1-2v)^{11}(1+2v) (\mathbf A_1 +\mathbf B_1)(\mathbf A_1 -\mathbf B_1)^2 (\mathbf A_1 +\mathbf B_1) \right) \\
&\quad \times
\det\!\left(I_{2^{0}} - u^{4}(1-2v)^{11}(1+2v)(\mathbf A_1 + \mathbf B_1)(\mathbf A_1 - \mathbf B_1)^3\right) \\
&\quad \times
\det\!\left(I_{2^{0}} - u^{4} (1-2v)^{9} (1+2v)^3(\mathbf A_1- \mathbf B_1)(\mathbf A_1 + \mathbf B_1)^{3}\right) .
\end{align*}
Since $\mathbf A_1 = 1$ and $\mathbf B_1 = 0$, we have
\begin{align*}
\det\!\left(I_{2^{4}} - u\,Q_{4}^{(g)}\right)
&= \bigl(1 - u^2 (1 + 2v)^2 (1 - 2v)^{4}\bigr) 
   \bigl(1 + u(1 + 2v)^{2}(2v - 1)\bigr)
   \bigl(1 - u(2v + 1)^{3}\bigr) \\
&\quad \times \!\left(1 - u^{4}(1-2v)^{11}(1+2v) \right)^2 (1-u^4 (1-2v)^{9} (1+2v)^3).
\end{align*} 
Hence, the result follows. 
\end{proof}

The following result holds for $\zeta(Q_f^{(l)}, \PM_5, u)$.

\begin{thm}[Case of $N=5$]

\begin{align*}
\zeta(Q_f^{(l)}, \PM_5, u)
&= (1 - u^2 (1+2v)^4 (1-2v)^4)^{-1/32}
(1 - u(2v+1)^3(1-2v))^{-1/32}
(1 - u(1+2v)^4)^{-1/32} \\
&\quad \times (1 - u^4(1+2v)^5 (1-2v)^{11} )^{-1/32}
(1 - u^4(1-2v)^{13}(1+2v)^3)^{-1/32} \\
&\quad \times (1 - u^4(1-2v)^{15}(1+2v))^{-1/32}
(1 - u^8(1-2v)^{23}(1+2v)^9)^{-1/32} \\
&\quad \times (1 - u^8(1-2v)^{29}(1+2v)^3)^{-1/32}.
\end{align*}
\end{thm}

\begin{proof}
By Corollary 5.1, we have
\begin{align*}
\det\bigl(I_{2^5} - u Q^{(g)}_5\bigr)
&= (1 - u^2 (1+2v)^4 (1-2v)^4 ) 
(1 - u(2v+1)^3(1-2v))
(1 - u(1+2v)^4) \\
&\quad \times 
\det\bigl(I_{2^2} - u(2v+1)(1-2v)(\mathbf A_3 - \mathbf B_3)\bigr) \\
&\quad \times 
\det\bigl(I_{2^3} - u(1-2v)(\mathbf A_4 - \mathbf B_4)\bigr)
\det\bigl(I_{2^4} - u(\mathbf A_5 -\mathbf B_5)\bigr).
\end{align*}

By Proposition 8.1 and the fact that $\mathbf A_1 = 1,\mathbf  B_1 = 0$, we have
\begin{align*}
\det\bigl(I_{2^2} - u(2v+1)(1-2v)(\mathbf A_3 - \mathbf B_3)\bigr)
&= \det\bigl(I_{2^0} - u^4(2v+1)^5 (1-2v)^{11} 
(\mathbf A_1 +\mathbf B_1)(\mathbf A_1 - \mathbf B_1)^2(\mathbf A_1 +\mathbf B_1)\bigr) \\
&=( 1 - u^4(2v+1)^5 (1-2v)^{11} ) .
\end{align*}

By Proposition 8.2, we get
\begin{align*}
\det\bigl(I_{2^3} - u(1-2v)(\mathbf A_4 -\mathbf B_4)\bigr)
&= \bigl(1 - u^4(1-2v)^4(1-2v)^9(1+2v)^3
(\mathbf A_1 -\mathbf B_1)(\mathbf A_1 +\mathbf B_1)^3\bigr) \\
&\quad \times
\bigl(1 - u^4(1-2v)^4(1-2v)^{11}(1+2v)
(\mathbf A_1 +\mathbf B_1)(\mathbf A_1 - \mathbf B_1)^3\bigr) \\
&= \bigl(1 - u^4(1-2v)^{13}(1+2v)^3\bigr)
\bigl(1 - u^4(1-2v)^{15}(1+2v)\bigr).
\end{align*}

Furthermore, by Proposition 8.2, we have
\begin{align*}
\det\bigl(I_{2^4} - u(\mathbf A_5 -\mathbf B_5)\bigr)
&= \det\bigl(I_2 - u^4(1-2v)^9(1+2v)^3
(\mathbf A_2 -\mathbf B_2)(\mathbf A_2 +\mathbf B_2)^3\bigr) \\
&\quad \times
\det\bigl(I_2 - u^4(1-2v)^{11}(1+2v)
(\mathbf A_2 +\mathbf B_2)(\mathbf A_2 -\mathbf B_2)^3\bigr).
\end{align*}

Since
\[
\mathbf A_2 +\mathbf B_2 =
\begin{bmatrix}
1 & 2v \\
2v & 1
\end{bmatrix},
\qquad
\mathbf A_2 - \mathbf B_2 = (1-2v)
\begin{bmatrix}
1 & 0 \\
0 & -1
\end{bmatrix},
\]
we have
\[
(\mathbf A_2 -\mathbf B_2)(\mathbf A_2 +\mathbf B_2)^3
= (1-2v)
\begin{bmatrix}
1 + 12v^2 & 6v + 8v^3 \\
-(6v + 8v^3) & -(1 + 12v^2)
\end{bmatrix}.
\]

Now, let $w = u^4(1-2v)^9(1+2v)^3$. Then we have
\begin{align*}
\det\bigl(I_2 - w(\mathbf A_2 -\mathbf B_2)(\mathbf A_2 +\mathbf B_2)^3\bigr)
&= \det
\begin{bmatrix}
1 - w(1-2v)(1+12v^2) & -w(1-2v)(6v+8v^3) \\
w(1-2v)(6v+8v^3) & 1 + w(1-2v)(1+12v^2)
\end{bmatrix}.
\end{align*}

Let $z = w(1-2v) = u^4(1-2v)^{10}(1+2v)^3$. Then we get
\begin{align*}
\det\bigl(I_2 - w(\mathbf A_2 -\mathbf B_2)(\mathbf A_2 +\mathbf B_2)^3\bigr)
&= \det
\begin{bmatrix}
1 - z(1+12v^2) & -z(6v+8v^3) \\
z(6v+8v^3) & 1 + z(1+12v^2)
\end{bmatrix} \\
&= (1 - z(1+12v^2))(1 + z(1+12v^2)) + z^2(6v+8v^3)^2 \\
&= 1 - z^2(1+2v)^3(1-2v)^3 \\
&= 1 - u^8(1-2v)^{23}(1+2v)^9 .
\end{align*}

Next, we have
\[
(\mathbf A_2 +\mathbf B_2)(\mathbf A_2 -\mathbf B_2)^3
= (1-2v)^3
\begin{bmatrix}
1 & -2v \\
2v & -1
\end{bmatrix}.
\]

Let $\tilde z = u^4(1-2v)^{14}(1+2v)$. Then we have
\begin{align*}
\det\bigl(I_2 - w(\mathbf A_2 +\mathbf B_2)(\mathbf A_2 -\mathbf B_2)^3\bigr)
&= \det
\begin{bmatrix}
1 - \tilde z & 2v\tilde z \\
-2v\tilde z & 1 + \tilde z
\end{bmatrix} \\
&= 1 - \tilde z^2 + 4v^2\tilde z^2
= 1 - \tilde z^2(1-4v^2) \\
&= 1 - u^8(1-2v)^{29}(1+2v)^3 .
\end{align*}
Therefore, the result follows. 
\end{proof}

\section{Zeta function of Rule 90 for general $N$}

From now on, we consider the ordinary Rule 90, which corresponds to the case \(v=0\)
of fuzzy Rule 90.  In this case, the local operator is denoted by
\(Q_{90}^{(l)}\).

In Komatsu et al. \cite{KKS3}, the dyadic product formula for the zeta function of Rule 90 was shown for \(N=2,3,4\), and was stated as a
conjecture for general \(N\).  In this section, we prove that this formula holds for
all \(N\geq2\). Therefore their conjecture is correct.

\begin{thm}[Dyadic product formula for the zeta function of Rule 90]
\label{thm:Rule90-dyadic}
For each \(m\in\mathbb Z_{\ge}\), let \(N\in\mathbb Z_{>}\) be determined by
\[
2^{m-1}<N\leq 2^m.
\]
That is,
\[
\begin{aligned}
m=1 &\quad\Longrightarrow\quad N=2,\\
m=2 &\quad\Longrightarrow\quad N\in\{3,4\},\\
m=3 &\quad\Longrightarrow\quad N\in\{5,6,7,8\},
\end{aligned}
\]
and so on.
Then, as a formal power series in \(u\),
\[
\zeta(Q_{90}^{(l)},\PM_N,u)^{-1}
=
(1-u^{2^m})^{2^{-m}}
\prod_{k=0}^{m-1}
\left(
\frac{1-u^{2^k}}{1+u^{2^k}}
\right)^{2^{-(N-2^k+k+1)}}.
\]
For \(N=1\), we have
\[
\zeta(Q_{90}^{(l)},\PM_1,u)^{-1}=1-u.
\]
\end{thm}

\begin{proof}
We identify a configuration
\[
\eta=(\eta_0,\eta_1,\ldots,\eta_{N-1})
\]
with an element of \(\mathbb F_2^N\), where $\mathbb{F}_p$ is the finite field of order $p$ associated with the prime $p$.  Under Rule 90, the local update is
\[
(i,j)\longmapsto (i+j,j),
\]
where the addition is taken in \(\mathbb F_2\).  Hence the global Rule 90 map is
the linear map
\[
T_N:\mathbb F_2^N\longrightarrow \mathbb F_2^N
\]
defined by
\[
T_N(\eta_0,\eta_1,\ldots,\eta_{N-1})^T
=
(\eta_0+\eta_1,\eta_1+\eta_2,\ldots,\eta_{N-2}+\eta_{N-1},\eta_{N-1})^T.
\]
Let
\[
U_N(\eta_0,\eta_1,\ldots,\eta_{N-1})^T
=
(\eta_1,\eta_2,\ldots,\eta_{N-1},0)^T.
\]
Then
\[
T_N=I_N+U_N,
\qquad
U_N^N=0.
\]
Since $U_N^N=0$ and the ground field is $\mathbb{F}_2$,
we have
\[
(I_N+U_N)
\left(I_N+U_N+U_N^2+\cdots+U_N^{N-1}\right)
=
I_N+U_N^N
=
I_N.
\]
The same identity holds with the two factors reversed.
Therefore $T_N$ is invertible, with
\[
T_N^{-1}
=
I_N+U_N+U_N^2+\cdots+U_N^{N-1}.
\]

Thus $T_N$ is a bijection of $\mathbb{F}_2^N$.
Consequently, the global matrix of Rule 90, $Q_{N,90}^{(g)}$ is the
permutation matrix induced by $T_N$.
Its $r$th power is the permutation matrix induced by $T_N^r$,
and hence
\[
\begin{aligned}
\operatorname{Tr}\!\left(\left(Q_{N,90}^{(g)}\right)^r\right)
&=\#\left\{
\eta\in\mathbb{F}_2^N : T_N^r\eta=\eta
\right\}
\\
&=\#\ker(T_N^r-I_N).
\end{aligned}
\]

For $r\in\mathbb{Z}_{>0}$, $r$ can be uniquely expressed as
\[
r=2^k(2s-1),
\]
with $k\in\mathbb{Z}_{\geq}$ and $s\in\mathbb{Z}_{>}$.
Since the characteristic of \(\mathbb F_2\) is \(2\), we have
\[
(I_N+U_N)^{2^k}=I_N+U_N^{2^k}.
\]
Therefore,
\[
T_N^r-I_N
=
(I_N+U_N)^{2^k(2s-1)}-I_N
=
(I_N+U_N^{2^k})^{2s-1}-I_N.
\]
The polynomial
\[
(1+x)^{2s-1}-1
\]
has constant term \(0\).  Hence it is divisible by \(x\).  Therefore,
there exists a polynomial \(R(x)\) such that
\[
(1+x)^{2s-1}-1=xR(x).
\]
Moreover, the coefficient of \(x\) in \((1+x)^{2s-1}-1\) is
\(2s-1\), which is equal to \(1\) in \(\mathbb F_2\).  Hence
\[
R(0)=1.
\]
Thus
\[
T_N^r-I_N
=
U_N^{2^k}R(U_N^{2^k}).
\]
Since \(R(0)=1\), the matrix
\(R(U_N^{2^k})\) is invertible.  Hence
\[
\ker(T_N^r-I_N)=\ker(U_N^{2^k}).
\]

Now \(U_N^{2^k}\) is the \(2^k\)-step shift.
For example, when \(N=3\), we have
\[
U_3=
\begin{bmatrix}
0&1&0\\
0&0&1\\
0&0&0
\end{bmatrix},
\qquad
U_3^2=
\begin{bmatrix}
0&0&1\\
0&0&0\\
0&0&0
\end{bmatrix}.
\]
Thus
\[
U_3
\begin{bmatrix}
x\\ y\\ z
\end{bmatrix}
=
\begin{bmatrix}
y\\ z\\ 0
\end{bmatrix}.
\]
Hence \(U_3(x,y,z)^T=0\) if and only if \(y=z=0\), and therefore
\[
\ker U_3
=
\left\{
\begin{bmatrix}
x\\0\\0
\end{bmatrix}
:x\in\mathbb F_2
\right\}.
\]
Thus
\[
\dim_{\mathbb F_2}(\ker U_3)=1.
\]
Similarly,
\[
U_3^2
\begin{bmatrix}
x\\ y\\ z
\end{bmatrix}
=
\begin{bmatrix}
z\\0\\0
\end{bmatrix}.
\]
Hence \(U_3^2(x,y,z)^T=0\) if and only if \(z=0\), and therefore
\[
\ker U_3^2
=
\left\{
\begin{bmatrix}
x\\y\\0
\end{bmatrix}
:x,y\in\mathbb F_2
\right\}.
\]
Thus
\[
\dim_{\mathbb F_2}(\ker U_3^2)=2.
\]
Moreover, since \(U_3^3=0\), we have
\[
\dim_{\mathbb F_2}(\ker U_3^3)=3.
\]

We now prove the following formula for every integer $\ell\geq 0$:
\begin{equation}
\dim_{\mathbb{F}_2}\bigl(\ker U_N^\ell\bigr)
=\min\{\ell,N\}.
\label{eq:kernel-dimension}
\end{equation}

If $\ell=0$, then $U_N^0=I_N$, so
\[
\dim_{\mathbb{F}_2}(\ker U_N^0)=0.
\]
If $1\leq\ell<N$, then
\[
U_N^\ell
\begin{bmatrix}
x_0\\
x_1\\
\vdots\\
x_{N-1}
\end{bmatrix}
=
\begin{bmatrix}
x_\ell\\
x_{\ell+1}\\
\vdots\\
x_{N-1}\\
\underbrace{0,\ldots,0}_{\ell\text{ entries}}
\end{bmatrix}.
\]
Thus $U_N^\ell (x_0,x_1,...,x_{N-1})^{T}=0$ if and only if
\[
x_\ell=x_{\ell+1}=\cdots=x_{N-1}=0.
\]
The first $\ell$ coordinates are arbitrary, and therefore
\[
\ker U_N^\ell
=
\left\{
(x_0,\ldots,x_{\ell-1},0,\ldots,0)^{\mathsf T}
:
x_0,\ldots,x_{\ell-1}\in\mathbb{F}_2
\right\}.
\]
Hence
\[
\dim_{\mathbb{F}_2}(\ker U_N^\ell)=\ell.
\]

If $\ell\geq N$, then $U_N^\ell=0$, since $U_N^N=0$.
Consequently,
\[
\ker U_N^\ell=\mathbb{F}_2^N,
\qquad
\dim_{\mathbb{F}_2}\ker U_N^\ell=N.
\]
Combining these cases proves the formula (10).
In particular,
\[
\dim_{\mathbb{F}_2}(\ker U_N^{2^k})
=
\min\{2^k,N\}.
\]
Together with
\[
\ker(T_N^r-I_N)=\ker U_N^{2^k},
\qquad
r=2^k(2s-1),
\]
this yields
\[
\begin{aligned}
\operatorname{Tr}\!\left(\left(Q_{N,90}^{(g)}\right)^r\right)
&=\#\ker(T_N^r-I_N)
\\
&=\#\ker\!\left(U_N^{2^k}\right)
\\
&=2^{\dim_{\mathbb{F}_2}\left(\ker (U_N^{2^k})\right)}
\\
&=2^{\min\{2^k,N\}}.
\end{aligned}
\]
For example, we consider the case $N=3$ and $r=1$.
Then $k=0$ and $s=1$ in the representation $r=2^k(2s-1)$.
In this case,
\[
Q_{3,90}^{(g)}
=
\begin{bmatrix}
1 & 0 & 0 & 0 & 0 & 0 & 0 & 0 \\
0 & 0 & 0 & 0 & 0 & 0 & 0 & 1 \\
0 & 0 & 0 & 0 & 0 & 0 & 1 & 0 \\
0 & 1 & 0 & 0 & 0 & 0 & 0 & 0 \\
0 & 0 & 0 & 0 & 1 & 0 & 0 & 0 \\
0 & 0 & 0 & 1 & 0 & 0 & 0 & 0 \\
0 & 0 & 1 & 0 & 0 & 0 & 0 & 0 \\
0 & 0 & 0 & 0 & 0 & 1 & 0 & 0
\end{bmatrix},
\]
and hence
\[
\operatorname{Tr}\!\left(Q_{3,90}^{(g)}\right)=2.
\]

Moreover, since $T_3-I_3=U_3$, we have
\[
\begin{aligned}
\ker(T_3-I_3)
&=\ker\!\left(U_3^{2^0}\right)
=\ker (U_3)
\\[4pt]
&=\left\{
\begin{bmatrix}
x\\0\\0
\end{bmatrix}
:x\in\mathbb{F}_2
\right\}
\\[4pt]
&=\left\{
\begin{bmatrix}
0\\0\\0
\end{bmatrix},
\begin{bmatrix}
1\\0\\0
\end{bmatrix}
\right\}.
\end{aligned}
\]
Therefore,
\[
\#\ker (U_3)=2,
\qquad
\dim_{\mathbb{F}_2}(\ker (U_3))
=1
=\min\{2^0,3\}.
\]
Consequently,
\[
\begin{aligned}
\operatorname{Tr}\!\left(Q_{3,90}^{(g)}\right)
&=\#\ker(T_3-I_3)
\\
&=\#\ker (U_3)
\\
&=2^{\dim_{\mathbb{F}_2}(\ker (U_3))}
\\
&=2^{\min\{2^0,3\}}
=2.
\end{aligned}
\]
This verifies the trace formula for $N=3$ and $r=1$.
Therefore, we obtain
\begin{equation}
\operatorname{Tr}\!\left(\left(Q_{N,90}^{(g)}\right)^r\right)
=2^{\min\{2^k,N\}}.
\label{eq:trace-formula}
\end{equation}

By the trace formula for the IPS zeta function,
\[
C_r(Q_{90}^{(l)},\PM_N)
=
\frac1{2^N}
\operatorname{Tr}\left((Q_{N,90}^{(g)})^r\right),
\]
we obtain
\[
C_{2^k(2s-1)}(Q_{90}^{(l)},\PM_N)
=
2^{\min\{2^k,N\}-N}.
\]

First, we fix $m\in\mathbb{Z}_{>}$.
Next, we consider an integer $N$ satisfying
\[
2^{m-1}<N\leq 2^m.
\]
Then, for \(0\leq k\leq m-1\), we have \(2^k<N\), while for \(k\geq m\),
we have \(2^k\geq N\).  Hence
\[
C_{2^k(2s-1)}(Q_{90}^{(l)},\PM_N)
=
\begin{cases}
2^{2^k-N}, & 0\leq k\leq m-1,\\[2mm]
1, & k\geq m.
\end{cases}
\]

Using
\[
\log\!\left\{
\zeta\!\left(Q_{90}^{(l)},\PM_N,u\right)^{-1}
\right\}
=
-\sum_{r=1}^{\infty}
C_r(Q_{90}^{(l)},\PM_N)\frac{u^r}{r},
\]
we get
\[
\begin{aligned}
\log\!\left\{
\zeta\!\left(Q_{90}^{(l)},\PM_N,u\right)^{-1}
\right\}
&=
-\sum_{k=0}^{m-1}
2^{2^k-N}
\sum_{s=1}^{\infty}
\frac{u^{2^k(2s-1)}}{2^k(2s-1)}
-
\sum_{q=1}^{\infty}
\frac{u^{2^m q}}{2^m q} \\[1mm]
&=
-\sum_{k=0}^{m-1}
2^{-(N-2^k+k)}
\tanh^{-1}(u^{2^k})
+
2^{-m}\log(1-u^{2^m}).
\end{aligned}
\]
Therefore, we obtain
\begin{equation}
\begin{aligned}
&\log\!\left\{
\zeta\!\left(Q_{90}^{(l)},\PM_N,u\right)^{-1}
\right\}
\\[4pt]
&=\frac{1}{2^m}\log\!\left(1-u^{2^m}\right)
-\sum_{k=0}^{m-1}
\frac{1}{2^{N-(2^k-k)}}
\tanh^{-1}\!\left(u^{2^k}\right).
\end{aligned}
\label{eq:zeta-rule90}
\end{equation}
This is exactly the formula given in Komatsu et al. \cite{KKS3} for $m=1,2$. Hence we have proved this formula for any 
$m\in\mathbb{Z}_{>}$.
Finally, combining (12) with the following relation
\[
\tanh^{-1}x
=
\frac12\log\left(\frac{1+x}{1-x}\right),
\]
we obtain
\[
\zeta(Q_{90}^{(l)},\PM_N,u)^{-1}
=
(1-u^{2^m})^{2^{-m}}
\prod_{k=0}^{m-1}
\left(
\frac{1-u^{2^k}}{1+u^{2^k}}
\right)^{2^{-(N-2^k+k+1)}}.
\]
This completes the proof.
\end{proof}

\section*{Acknowledgments}
The fourth author is partially supported by JSPS KAKENHI
Grant-in-Aid for Scientific Research (C)
(Grant No.~JP26K06897).

\end{document}